\documentclass[twocolumn, aps, unsortedaddress]{revtex4-2}
\usepackage{graphicx}
\usepackage{amsmath}
\usepackage{graphics}
\usepackage{epsfig}

\usepackage{amsmath}
\usepackage{amsfonts}
\usepackage{amssymb}
\usepackage{xcolor}
\usepackage{subcaption}
\usepackage{float}
\DeclareUnicodeCharacter{2212}{-}

\usepackage[normalem]{ulem}

\begin{document}

\newcommand{\lyxdot}{.}

\title{ Current-driven mechanical motion of double stranded DNA results in
structural instabilities and chiral-induced-spin-selectivity of electron
transport}
\author{Nicholas S. Davis}
\affiliation{College of Science and Engineering, James Cook University, Townsville,
QLD, 4811, Australia}
\author{Julian A. Lawn}
\affiliation{College of Science and Engineering, James Cook University, Townsville,
QLD, 4811, Australia}
\author{Riley J. Preston}
\affiliation{Institute of Physics, University of Freiburg, Hermann-Herder-Strasse
3, 79104 Freiburg, Germany}
\author{Daniel S. Kosov}
\affiliation{College of Science and Engineering, James Cook University, Townsville,
QLD, 4811, Australia}
\begin{abstract}
Chiral-induced-spin-selectivity of electron transport and its interplay
with DNA's mechanical motion is explored in a double stranded DNA
helix with spin-orbit-coupling. The mechanical degree of freedom is
treated as a stochastic classical variable experiencing fluctuations
and dissipation induced by the environment as well as force exerted
by nonequilibrium, current-carrying electrons. Electronic degrees
of freedom are described quantum mechanically using nonequilibrium
Green's functions. Nonequilibrium Green's functions are computed along
the trajectory for the classical variable taking into account dynamical,
velocity dependent corrections. This mixed quantum-classical approach
enables calculations of time-dependent spin-resolved currents. We
showed that the electronic force may significantly modify the classical
potential which, at sufficient voltage, creates a bi-stable potential
with considerable effect on electronic transport. The DNA's mechanical
motion has a profound effect on spin transport; it results in chiral-induced
spin selectivity increasing spin polarisation of the current by 9\%
and also resulting in temperature-dependent current voltage characteristics.
We demonstrate that the current noise measurement provides an accessible
experimental means to monitor the emergence of mechanical instability
in DNA motion. The spin resolved current noise also provides important
dynamical information about the interplay between vibrational and spin
degrees of freedom in DNA. 
\end{abstract}
\maketitle

\section{Introduction}

Chiral-induced-spin-selectivity (CISS) is an intriguing phenomenon
observed in a wide range of molecules and conditions \cite{aiello2022,naaman2019a,naaman2020b,naaman2015}.
Primarily this effect pertains to the preferential spin current flowing
in each direction across the molecule, attracting interest in biochemistry
due to the link between spin currents and enantioselectivity \cite{michaeli2016}
and controlling reactions \cite{naaman2020,evers2022} as well as
spintronics \cite{naaman2012,bloom2017,mondal2016,yang2019,shang2022,al2018}
due to the magnitude of spin-polarisation (SP) \cite{kulkarni2020}.
This significant SP is surprising given the weak spin-orbit-coupling
(SOC) in organic molecules\cite{hongki2006}, and many avenues have
been explored to reconcile the difference between experimental observation
and theoretical prediction. Consequently, the emergence of a large
SP has been attributed to many different effects, yet a definitive
description of CISS remains elusive.

Measurements of the CISS effect have continued since its discovery
in attempts to elucidate its underlying mechanisms. This has lead to
a few key observations that warrant consideration when designing effective
models. The first is that the magnitude of SP increases with the length
of the molecule\cite{fransson2021,mishra2020,kettner2018}, indicating
a compounding mechanism as electrons traverse the molecule. Also apparent
is the sensitivity of SP to temperature\cite{fransson2023a}; sparking
a strong interest in vibrational effects.

It has been well established that there is no spin separability in
a system involving only a single transport channel. Time-reversibility
of SOC allows its removal via a gauge transformation; thus spin cannot
separate\cite{yang2019,bardarson2008,vittmann2022}. In junctions
with single-point contact to only two leads, it is necessary for the
molecule itself to possess multiple spin-transport channels, which
is the case for double-stranded DNA with inter-strand exchange.
  {Furthermore, it has been demonstrated that while a two-terminal setup using a ferromagnet is sufficient to detect CISS effects in a non-linear transport regime \cite{yang2020}, a multi-terminal setup is necessary for the linear response  regime \cite{yang2019}}.

Beyond attempts to magnify the SOC within the molecule \cite{dalum2019,gutierrez2012,yeganeh2009},
several aspects of molecular transport have featured prominently as
candidates for core functions of CISS. Many models neglect electron-electron
interactions, yet their inclusion has proved to greatly enhance SP
\cite{fransson2019,kumar2022}. A similar case holds for electron-phonon
interactions when including nuclear vibrations \cite{diaz2018a,fransson2021}.
Furthermore, these effects are necessary to include as the single-electron
treatment cannot fully describe CISS \cite{fransson2022c,fransson2022b}.

The importance of nuclear vibrations to CISS is expanded upon here, where we consider DNA's mechanical
motion coupled to electronic degrees of freedom. The mechanical motion
is considered classically, in which it evolves in time under the influence
of the force produced by nonequilibrium quantum electrons. Inherently
advantageous to this approach is the versatility of the mechanical
coordinate, which can experience large amplitude motion in complex
potential energy landscapes. The dynamics of DNA's mechanical motion
is modelled using a Langevin equation with nonequilibrium electronic
forces computed via nonequilibrium Green's functions (NEGF). These
calculations of time-dependent current include non-adiabatic dynamical
corrections arising from the time-dependence of the electronic Hamiltonian
which depends parametrically on the mechanical variable. Aside from
the SP, which is the topic of most studies, this approach facilitates
the calculation of spin-resolved current noise, where dynamical corrections   {    arising due to a breakdown of the Born-Oppenheimer approximation}
play an important role.

The paper is organised as follows. Section II describes the theory;
the model Hamiltonian, NEGF calculations of current and forces, and
the numerical scheme to solve the Langevin equation with NEGF forces.
Results are presented and discussed in Section III, focusing on
the average nonequilibrium energy landscape for DNA's mechanical motion,
along with current and current noise induced by DNA's mechanical motion.
Section IV presents the conclusions.

We use atomic units in our equations: $\hbar=e=1$. Most values of
physical quantities will also be stated in atomic units (a.u.); however,
we will present the values for all energy related quantities in eV
and voltages in V, for clarity.

\section{Theory}

\subsection{Model}

\begin{figure}
\begin{centering}
\includegraphics[scale=0.12]{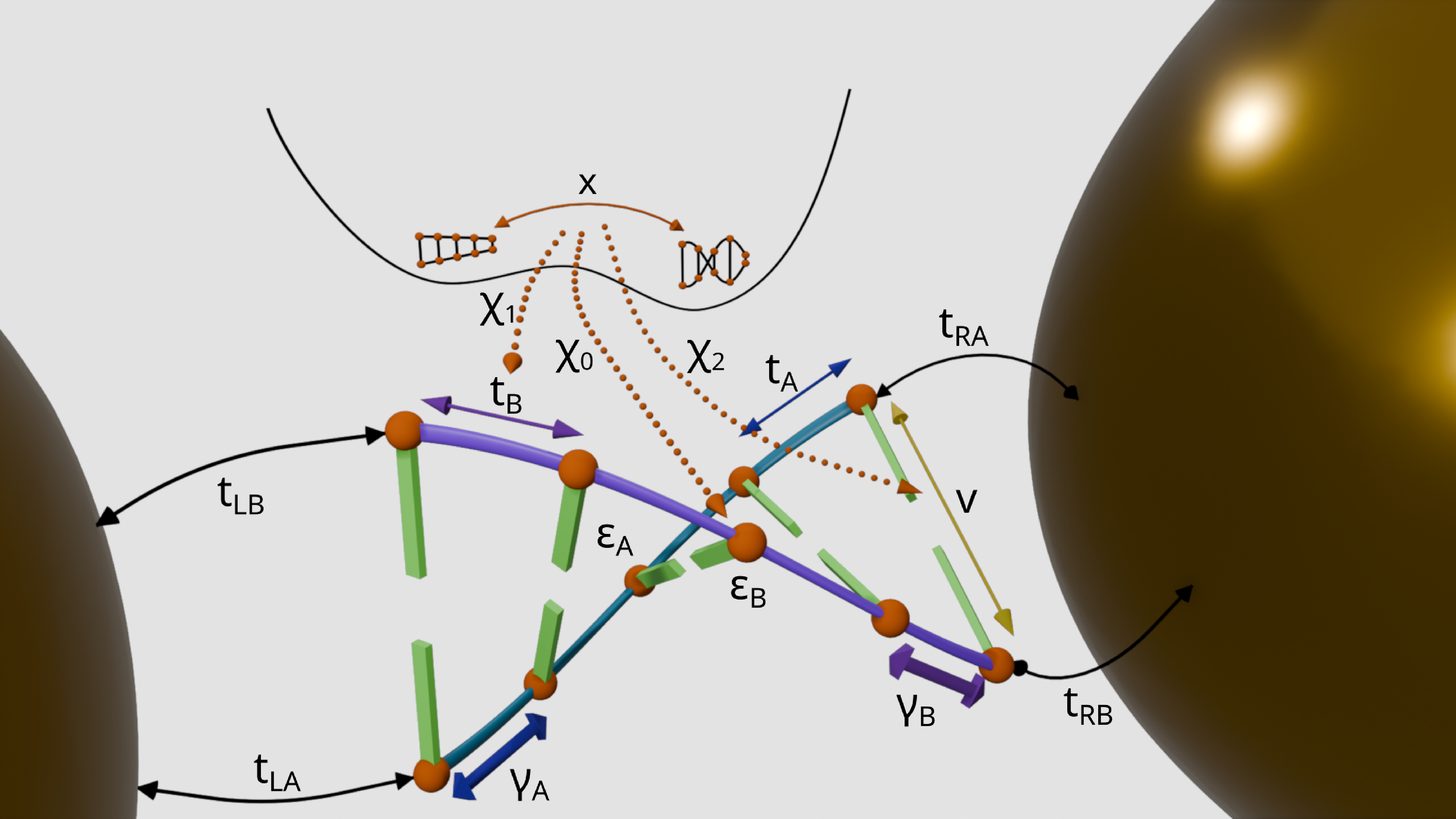} 
\par\end{centering}
\caption{
  {Schematic of the model system.
Double stranded DNA connects two electrodes. The DNA's mechanical motion influences the on-site energies,  intra-strand electronic hopping, and  inter-strand electron transfer.
$\gamma_A$ and $\gamma_B$ denote SOC, which is included in our model  as spin-dependent intra-strand coupling. Both strands of DNA are attached to the electrodes.} }
\label{Sketch} 
\end{figure}

The system consists of a double-stranded DNA molecule with the ends
linked to two macroscopic electrodes. Each nucleotide in the DNA is
represented as a single electronic level. The electrons are allowed
to hop within a single strand between neighbouring nucleotides as
well as between nucleotides within the base pair. The electronic Hamiltonian
depends on the DNA conformation, for which the nuclear dynamics are collectively described
by a single mechanical degree of freedom.

The system is described by the Hamiltonian 
\begin{multline}
H(t)=H_{M}+H_{L}+H_{R}+H_{ML}+H_{MR}\\
+H_{\text{e-mech}}(x)+H_{\text{mech}}(p,x),\label{1}
\end{multline}
which is partitioned into the DNA Hamiltonian $H_{M}$, left and right electrodes
$H_{L}+H_{R}$, the electronic coupling of the DNA to the electrodes $H_{ML}+H_{MR}$, the coupling
of tunneling electrons to the mechanical degree of freedom $H_{\text{e-mech}}$,
and, finally, the mechanical degree of freedom $H_{\text{mech}}(p,x)$.
  {The electronic Hamiltonian is time-dependent through the dependence on the mechanical degree of freedom $x(t)$.}
The electrodes' Hamiltonians take the form of non-interacting electronic
reservoirs: 
\begin{equation}
H_{L}+H_{R}=\sum_{\alpha k\sigma}\epsilon_{\alpha k\sigma}d_{\alpha k\sigma}^{\dagger}d_{\alpha k\sigma},
\end{equation}
where $d^{\dagger}$ and $d$ are the standard creation and annihilation
operators and the subscript $\alpha k\sigma$ denotes the action of the operator in
electrode $\alpha\in\left\{ L,R\right\} $ on single-particle state
$k$ with spin $\sigma\in\left\{ \uparrow,\downarrow\right\} $.
  {
A visualisation of our proposed molecular junction configuration is shown in Fig. 1.}

The molecular Hamiltonian describes a double-stranded DNA helix with
spin-orbit-coupling, as in {   Refs.} \cite{guo2012,du2020}: 
\begin{align}
H_{M} & =\sum_{\beta j\sigma}\epsilon_{\beta}d_{\beta j\sigma}^{\dagger}d_{\beta j\sigma}+\sum_{\beta j\sigma}t_{\beta}(d_{\beta j\sigma}^{\dagger}d_{\beta,j+1,\sigma}+\text{h.c.})\nonumber \\
 & +\sum_{j\sigma}v(d_{Aj\sigma}^{\dagger}d_{Bj\sigma}+\text{h.c.})+V_{\text{SOC}}.\label{eq:Hmol}
\end{align}
Here, subscript $j$ is the base-pair index for strand $\beta\in\left\{ A,B\right\} $,
$\epsilon_{\beta}$ is the on-site energy, while $t_{\beta}$ and $v$ are
the intra-chain and inter-chain hopping matrix elements, respectively.
These hopping integrals are treated as equal for all sites within
each strand. However, the strands are assumed to be different. As
such, adopting parameters similar to Refs. \cite{du2020}: $\epsilon_{A}=-0.2$ eV, $\epsilon_{B}=0.1$ eV,
$t_{A}=0.1$ eV, $t_{B}=-0.14$ eV and $v=-0.08$ eV.

The spin-orbit coupling (SOC) is 
\begin{equation}
V_{\text{SOC}}=\sum_{\beta j\sigma\sigma'}i\gamma_{\beta}\Lambda_{\sigma\sigma'}^{\beta j}d_{\beta j\sigma}^{\dagger}d_{\beta,j+1,\sigma'}+\text{h.c.}
\end{equation}
The SOC matrix $\Lambda_{\sigma\sigma'}^{\beta j}$ contains constants
determined by the geometry of the helix, defined by  helix angle $\theta$
and rotation per base pair $\phi$, 
\begin{align}
\Lambda_{\uparrow\uparrow}^{\beta j} & =2\cos\theta,\;\;\;\Lambda_{\downarrow\downarrow}^{\beta j}=-2\cos\theta,\\
\Lambda_{\uparrow\downarrow}^{\beta j} & =i(-)_{\beta}(e^{-i\phi(j-1)}+e^{-i\phi j})\sin\theta,\;\;\;\Lambda_{\downarrow\uparrow}^{\beta j}=\Big(\Lambda_{\uparrow\downarrow}^{\beta j}\Big)^{*},
\end{align}
where $(-)_{\beta}$ is (+1) if $\beta=A$ and (-1) if $\beta=B$.
The SOC strength constants are taken as $\gamma_{A}=-0.014$ eV and
$\gamma_{B}=0.01$ eV, similar to \cite{du2020}.

The mechanical degree of freedom $(x,p)$ is described by a Hamiltonian
for a classical particle of mass $m$ 
\begin{equation}
H_{\text{mech}}(x)=\frac{p^{2}}{2m}+U(x),
\end{equation}
with harmonic potential $U(x)=\frac{1}{2}m\omega_{0}x^{2}$. This
frequency is chosen within a range to reflect the vibrational dynamics
of DNA, with standard value $\omega_{0}=5$ meV \cite{lee2006}.
Finally, the mass, $m=5.9244\times10^{5}$ a.u., is taken as the approximate
mass of a single nucleotide. 
  {Quantum effects become important, and the classical treatment of nuclear dynamics generally fails, when the temperature is smaller than the characteristic frequency of nuclear motion. In this paper, we perform our calculations at $T=300$ K  (25.8 meV) and $T=77$ K (6.6 meV), where both temperatures are above the frequency of the DNA's mechanical motion, $\omega_0$,  considered in our model. Therefore, we expect that our classical approach provides a qualitatively correct description  given our focus on  the low energy mechanical motion of DNA and relatively high temperatures.}

The interaction between electrons and the mechanical degree of freedom
is assumed to be described by linear response to the classical coordinate
$x$, and is given by 
\begin{multline}
H_{\text{e-mech}}(x)=-\chi_{0}\sqrt{2m\omega_{0}}x\sum_{\beta j\sigma}d_{\beta j\sigma}^{\dag}d_{\beta j\sigma}\\
-\chi_{1}\sqrt{2m\omega_{0}}x\sum_{\beta j\sigma}(d_{\beta j\sigma}^{\dag}d_{\beta,j+1,\sigma}+\text{h.c.})\\
-\chi_{2}\sqrt{2m\omega_{0}}x\sum_{j\sigma}(d_{Aj\sigma}^{\dag}d_{Bj\sigma}+\text{h.c}).
\end{multline}
Here, the first term describes the on-site electron-mechanical motion
interaction, the second term takes into account the influence of the
mechanical motion on the intra-strand electronic hopping, and the last
term deals with inter-strand electron transfer assisted by the mechanical
motion of DNA. The corresponding coupling strengths are described
by $\chi_{0}$, $\chi_{1}$ and $\chi_{2}$. Following Refs. \cite{du2020,voityuk2001,senthilkumar2005},
we assume that $\chi_{1}=0.2\chi_{0}$, $\chi_{2}=\chi_{1}$. The
value of $\chi_{0}$ will be treated as a parameter to study the role
of the strength of the electron-vibration coupling on the results.

The system-electrode coupling is described by 
\begin{equation}
H_{ML}= \sum_{k\beta\sigma}\left(t_{Lk,\beta1}d_{Lk\sigma}^{\dagger}d_{\beta1\sigma}+\text{h.c}\right),
\end{equation}
\begin{equation}
H_{MR}=\sum_{k\beta\sigma}\left(t_{Rk,\beta N}d_{Rk\sigma}^{\dagger}d_{\beta N\sigma}+\text{h.c}\right),
\end{equation}
where $t_{\alpha k,\beta1}$ and $t_{\alpha k,\beta N}$
are the tunnelling amplitudes between electrodes'
states $\alpha k\sigma$ and corresponding DNA base-pair 1 or N for
left and right couplings, respectively. Thus, it is assumed that both
strands of DNA are attached to the electrodes.

\subsection{Green's Functions with nonadiabatic corrections \label{sec:Green's-Functions}}

The self-energies and Green's functions are matrices in molecular
spin-orbital space with dimension $(2\times2\times N)$, where $N$
is the number of base-pairs in the DNA, 2 is for the number of DNA
strands and another 2 for the spin degrees of freedom.

We treat self-energies in the wide-band approximation, within which the
matrix elements of the retarded component are 
\begin{align}
{\Sigma}_{L,\beta j\sigma,\beta'j'\sigma'}^{R} & =-\frac{i}{2}\Gamma_{L}\delta_{\beta\beta'}\delta_{j1}\delta_{j'1}\delta_{\sigma\sigma'},
\end{align}
\begin{align}
{\Sigma}_{R,\beta j\sigma,\beta'j'\sigma'}^{R} & =-\frac{i}{2}\Gamma_{R}\delta_{\beta\beta'}\delta_{jN}\delta_{j'N}\delta_{\sigma\sigma'},
\end{align}
where $\Gamma_{\alpha}$ is the level broadening function due to the
coupling to electrode $\alpha$. The advanced self energy matrices
are obtained via the Hermitian conjugate of the retarded components
\begin{equation}
\mathbf{\Sigma}_{\alpha}^{A}=(\mathbf{\Sigma}_{\alpha}^{R})^{\dagger}.
\end{equation}
Finally, the lesser self-energies are computed via the fluctuation-dissipation
relation as

\begin{equation}
\mathbf{\Sigma}_{\alpha}^{<}(\omega)=f_{\alpha}(\omega)(\mathbf{\Sigma}_{\alpha}^{A}-\mathbf{\Sigma}_{\alpha}^{R}),
\end{equation}
where 
\begin{equation}
f_{\alpha}(\omega)=\frac{1}{1+e^{(\omega-\mu_{\alpha})/k_{B}T}},
\end{equation}
is the Fermi-Dirac occupation number for electrode $\alpha$ with
chemical potential $\mu_{\alpha}$ and temperature $T$. In all calculations
we assume the the voltage $V$ is applied symmetrically: $\mu_{L}=V/2$
and $\mu_{R}=-V/2$.

Suppose we know a trajectory $(x(t),p(t))$ for the mechanical degree
of freedom. The electronic part of the Hamiltonian {   in Eq.} (\ref{1}) becomes
explicitly time-dependent through the parametric dependence on $x(t)$,
which means that the Green's functions should be obtained from the
solution of the full Keldysh-Kadanoff-Baym equations. Following {   Refs.} \cite{Kershaw17,Kershaw18,Kershaw19},
we use a time-separation technique to solve the Keldysh-Kadanoff-Baym
equations in the Wigner space, resulting in the expression 
\begin{equation}
\mathbf{G}(t,\omega)=\mathbf{G}_{(0)}(t,\omega)+\mathbf{G}_{(1)}(t,\omega).
\end{equation}
Here ,
\begin{equation}
\mathbf{G}(t,\omega)=\int d(t_{1}-t_{2})e^{i\omega(t_{1}-t_{2})}\mathbf{G}(t_{1},t_{2}),
\end{equation}
is the Wigner space Green's function and $t=(t_{1}+t_{2})/2$ is the
central time. $\mathbf{G}_{(0)}(t,\omega)$ is the adiabatic Green's
function (Green's function which instantaneously follows the changes
in the reaction coordinate) and $\mathbf{G}_{(1)}(t,\omega)$ contains
non-adiabatic corrections (accounting for the dynamics of mechanical
motion). The adiabatic Green's functions are the standard Green's functions
computed for a static Hamiltonian. The advanced and retarded Green's
functions are computed by matrix inversion 
\begin{equation}
\mathbf{G}_{(0)}^{A/R}(t,\omega)=\left(\omega-\mathbf{h}\left(x(t)\right)-\mathbf{\Sigma}^{A/R}\right)^{-1},
\end{equation}
where the matrix $\mathbf{h}$ is formed from the electronic Hamiltonian
including the coupling to the  mechanical motion term: 
\begin{align}
 & h_{\beta j\sigma,\beta'j'\sigma'}=(\epsilon_{\beta}-\chi_{0}\sqrt{2m\omega_{0}}x)\delta_{\beta\beta'}\delta_{jj'}\delta_{\sigma\sigma'}\nonumber \\
 & +(t_{\beta}-\chi_{1}\sqrt{2m\omega_{0}}x)\delta_{\beta\beta'}\delta_{j\pm1,j'}\delta_{\sigma\sigma'}\nonumber \\
 & +(v-\chi_{2}\sqrt{2m\omega_{0}}x)\left(1-\delta_{\beta\beta'}\right)\delta_{jj'}\delta_{\sigma\sigma'}\nonumber \\
 & +i\Lambda_{\sigma\sigma'}^{\beta j}\gamma_{\beta}\delta_{\beta\beta'}\delta_{j\pm1,j'}.
 \label{h-matrix}
\end{align}
The lesser adiabatic
Green's function is computed using advanced and retarded components
\begin{equation}
\mathbf{G}_{(0)}^{<}(t,\omega)=\mathbf{G}_{(0)}^{R}(t,\omega)\mathbf{\Sigma}^{<}(\omega)\mathbf{G}_{(0)}^{A}(t,\omega).\label{eq:GL0}
\end{equation}

The first order nonadiabatic corrections to the advanced and retarded
Green's functions are (functional dependence on $t$ and $\omega$
is suppressed for brevity) 
\begin{equation}
\mathbf{G}_{(1)}^{A/R}=\frac{1}{2i}\mathbf{G}_{(0)}^{A/R}\left[\mathbf{G}_{(0)}^{A/R},\dot{\mathbf{h}}\right]_{-}\mathbf{G}_{(0)}^{A/R},
\end{equation}
where $\dot{\mathbf{h}}$ is the time derivative of matrix (\ref{h-matrix})

\begin{align}
 & \dot{h}_{\beta j\sigma,\beta'j'\sigma'}= - \dot{x}\sqrt{2m\omega_{0}} \delta_{\sigma\sigma'}\Big(\chi_{0}\delta_{\beta\beta'}\delta_{jj'} \nonumber \\
 & +\chi_{1}\delta_{\beta\beta'}\delta_{j\pm1,j'} 
 +\chi_{2}\left(1-\delta_{\beta\beta'}\right)\delta_{jj'} \Big).
\end{align}
 The nonadiabatic correction to the lesser component
is 
\begin{multline}
\mathbf{G}_{(1)}^{<}=\mathbf{G}_{(0)}^{A/R}\mathbf{\Sigma}^{<}\mathbf{G}_{(1)}^{A}+\mathbf{G}_{(1)}^{R}\mathbf{\Sigma}^{<}\mathbf{G}_{(0)}^{A}\\
+\frac{1}{2i}\mathbf{G}_{(0)}^{R}\left(\dot{\mathbf{h}}\ \mathbf{G}_{(0)}^{R}\partial_{\omega}\mathbf{\Sigma}^{<}+\mathbf{G}_{(0)}^{<}\dot{\mathbf{h}}+\text{h.c}\right)\mathbf{G}_{(0)}^{A}.
\end{multline}

\subsection{Spin-resolved current with dynamical corrections and force exerted
by electrons on mechanical degree of freedom}

Spin resolved electronic current, which includes dynamical corrections
due to the mechanical motion, is obtained as in {   Refs.} \cite{Kershaw17,Kershaw18,Kershaw19,Kershaw2020}.
The adiabatic current (depends on instantaneous position $x(t)$) is given
by the standard expression 
\begin{multline}
J_{\alpha\sigma}^{(0)}(t)=\intop_{-\infty}^{\infty}\frac{d\omega}{2\pi}\text{Tr}\Big[\mathbf{G}_{(0)}^{<}\mathbf{\Sigma}_{\alpha\sigma}^{A}+\mathbf{G}_{(0)}^{R}\mathbf{\Sigma}_{\alpha\sigma}^{<}\\
-\mathbf{\Sigma}_{\alpha\sigma}^{<}\mathbf{G}_{(0)}^{A}-\mathbf{\Sigma}_{\alpha\sigma}^{R}\mathbf{G}_{(0)}^{<}\Big],\label{j0}
\end{multline}
while the first order correction is given by 
\begin{multline}
J_{\alpha\sigma}^{(1)}(t)=\intop_{-\infty}^{\infty}\frac{d\omega}{2\pi}\ \text{Tr}\Big[\mathbf{G}_{(1)}^{<}\mathbf{\Sigma}_{\alpha\sigma}^{A}\\
+\mathbf{G}_{(1)}^{R}\mathbf{\Sigma}_{\alpha\sigma}^{<}-\mathbf{\Sigma}_{\alpha\sigma}^{<}\mathbf{G}_{(1)}^{A}-\mathbf{\Sigma}_{\alpha\sigma}^{R}\mathbf{G}_{(1)}^{<}\\
+\frac{1}{2i}\left(\mathbf{\dot{G}}_{(0)}^{R}\partial_{\omega}\mathbf{\Sigma}_{\alpha\sigma}^{<}+\partial_{\omega}\mathbf{\Sigma}_{\alpha\sigma}^{<}\mathbf{\dot{G}}_{(0)}^{A}\right)\Big],\label{j1}
\end{multline}
where 
\begin{equation}
\mathbf{\dot{G}}_{(0)}^{A}=\mathbf{G}_{(0)}^{A}\mathbf{\dot{h}}\mathbf{G}_{(0)}^{A},
\end{equation}
\begin{equation}
\mathbf{\dot{G}}_{(0)}^{R}=\Big(\mathbf{\dot{G}}_{(0)}^{A}\Big)^{\dag},
\end{equation}
and the derivative $\partial_{\omega}\mathbf{\Sigma}_{\alpha\sigma}^{<}$
can be easily computed using the wide-band approximation expression
for $\mathbf{\Sigma}_{\alpha\sigma}^{<}$.

We turn our attention now to calculations of the force exerted by
nonequilibrium electrons on the mechanical degree of freedom. The
force operator is computed by differentiating the corresponding part
of the electronic Hamiltonian 
\begin{align}
 & \hat{F}_{\text{e}}(t)=-\partial_{x}H_{\text{e-mech}}(t)\nonumber \\
= & -\sum_{\beta j\sigma,\beta'j'\sigma'}(\partial_{x}h)_{\beta j\sigma,\beta'j'\sigma'}d_{\beta j\sigma}^{\dag}d_{\beta'j'\sigma'},
\end{align}
where 
\begin{multline}
(\partial_{x}h)_{\beta j\sigma,\beta'j'\sigma'}=- \sqrt{2m\omega_{0}} \delta_{\sigma\sigma'} \Big(\chi_{0}\delta_{\beta\beta'}\delta_{jj'}\\
+\chi_{1}\delta_{\beta\beta'}\delta_{j\pm1,j'}+\chi_{2}\left(1-\delta_{\beta\beta'}\right)\delta_{jj'} \Big).
\end{multline}
Taking a quantum average along with truncation
due to {   a time-scale} separation yields 
\begin{equation}
F_{\text{e}}(t)=i\int\frac{d\omega}{2\pi}\text{Tr}\left[\partial_{x}\mathbf{h}\mathbf{G}_{(0)}^{<}(t,\omega)\right].\label{fe}
\end{equation}
This force can then be integrated over $x$ and {    added to the classical potential} to produce an effective potential
for the classical coordinate produced by nonequilibrium current-carrying
electrons.

\subsection{Dynamics of {   the} mechanical degree freedom}

The mechanical degree of freedom $x$ is treated as a stochastic classical
variable experiencing fluctuations and dissipation induced by the
environment as well as force exerted by nonequilibrium, current carrying
electrons. To simulate the effect of the environment, the equation
of motion of the mechanical coordinate is defined by a Langevin equation,
\begin{equation}
p=m\dot{x},\label{le1}
\end{equation}
\begin{equation}
\dot{p}(t)=f(t)-\zeta(t)p+\eta(t),\label{le2}
\end{equation}
where the force 
\begin{equation}
f(t)=-m\omega_{0}^{2}x+F_{\text{e,neq}}(t),
\end{equation}
contains a classical oscillator term with spring constant $m\omega_{0}^{2}$,
and a nonequilibrium electronic force $F_{\text{e,neq}}$. The nonequilibrium
electronic force is defined as 
\begin{equation}
F_{\text{e,neq}}=F_{\text{e}}(x)-F_{\text{e,eq}}(x),
\end{equation}
where the electronic force $F_{\text{e}}$ is given by (\ref{fe})
and $F_{\text{e,eq}}(x)$ is the equilibrium electronic force computed
at zero voltage bias. This equilibrium force removal is justified
based on the assumption that the equilibrium potential energy surface
is completely represented by the harmonic potential term $\frac{1}{2}m\omega_{0}^{2}x^{2}$.
The environment terms, viscosity $\zeta$ and white noise random force
$\eta(t)$ are related to each other by the fluctuation dissipation
theorem $\langle\eta(t)\eta(t')\rangle=m\zeta kT\delta(t-t')$.

To integrate the Langevin equation (\ref{le1},\ref{le2}) we use
the BAOAB splitting algorithm\cite{leimkuhler2013}. The algorithm
breaks up the Langevin equation into three separate finite difference
equations: Inertial momentum update (B), position update (A) and a
momentum update due to the stochastic force (O); it involves NEGF
calculations of force on each internal momentum update step (B). This
algorithm has been shown to be more accurate in comparison to other
Langevin integrators, particularly being very robust to changes in
time-step; it also provides accurate barrier crossing times for 1D
double-welled potentials \cite{leimkuhler2013} which is important
for bistable DNA's mechanical motion considered in our paper. We use
integration time-step 1 a.u. and viscosity $\zeta=2.228\times10^{-4}$
a.u. in all our simulations.

\section{Results}

\subsection{Electronic current induces mechanical instability}

\begin{figure*}
\begin{centering}
\includegraphics[scale=0.6]{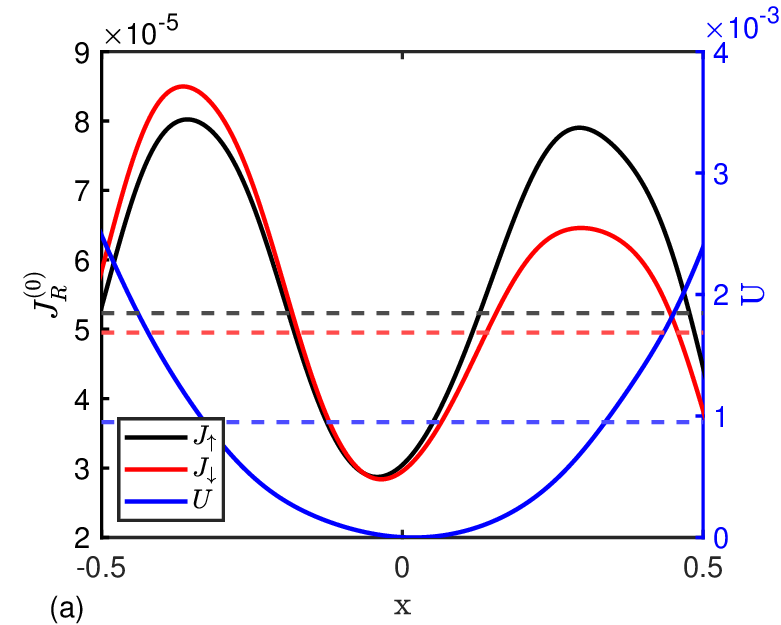} \includegraphics[scale=0.6]{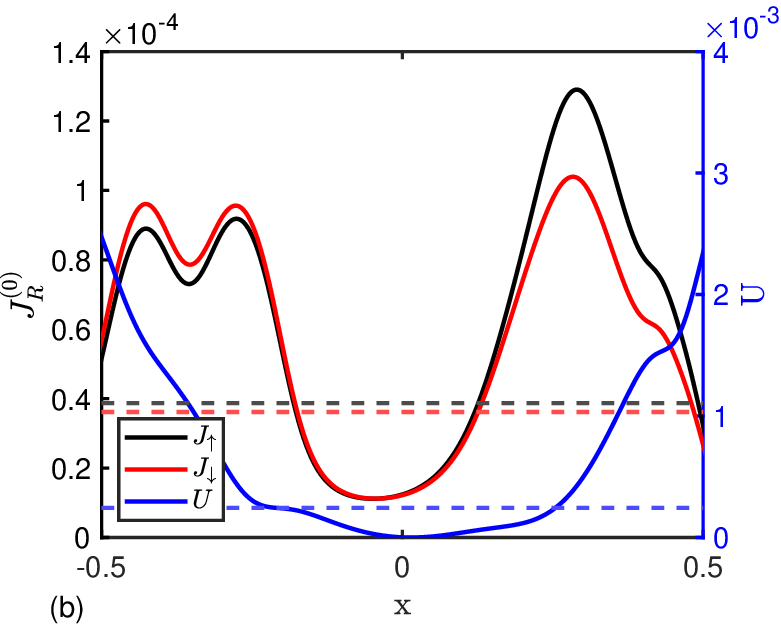} 
\par\end{centering}
\begin{centering}
\includegraphics[scale=0.6]{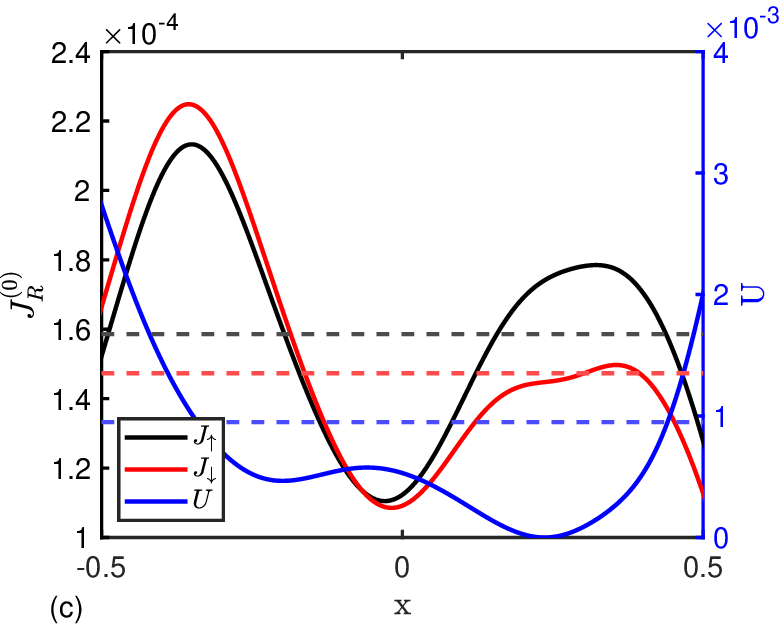} \includegraphics[scale=0.6]{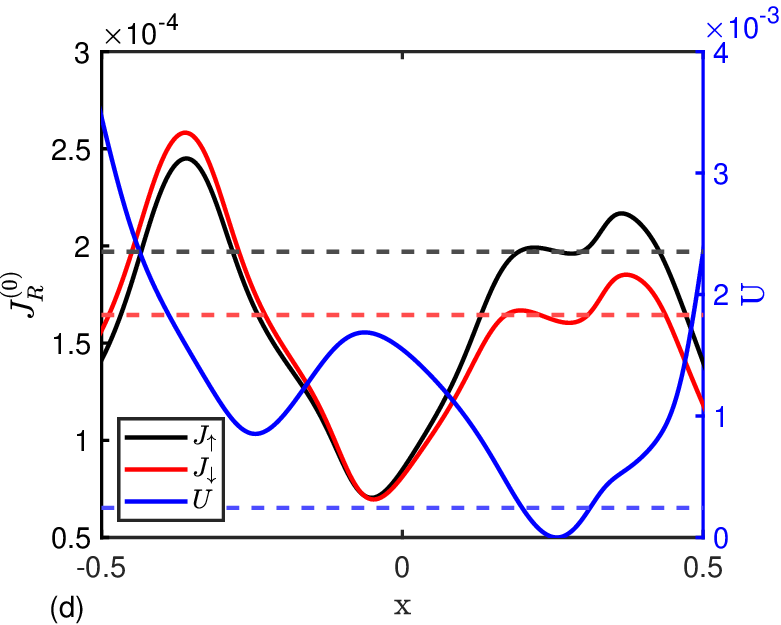} 
\par\end{centering}
\begin{centering}
\includegraphics[scale=0.6]{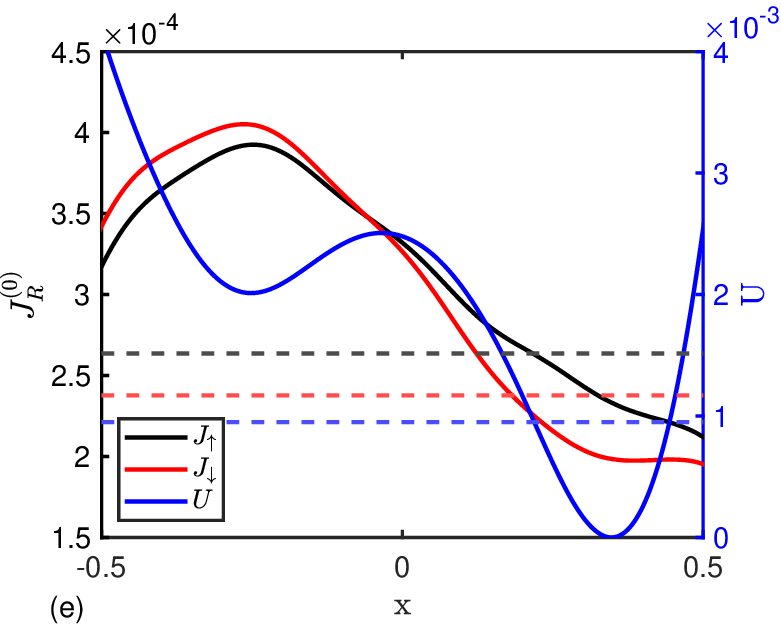} \includegraphics[scale=0.6]{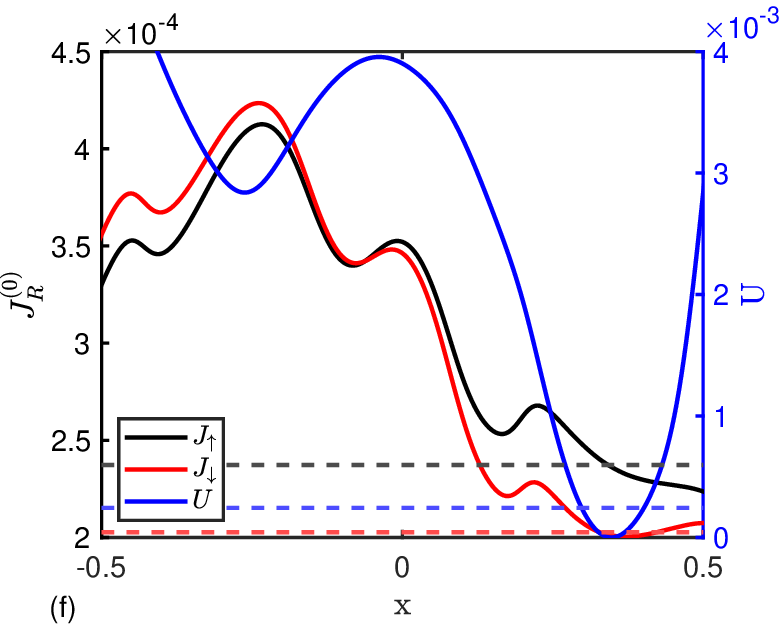} 
\par\end{centering}
\caption{Left-side axis: Spin-resolved current profile for adiabatic current
$J_{R\sigma}^{(0)}(x)$. Dashed horizontal lines indicate mean current
values computed via averaging over values of the current over Langevin
trajectory. Right-side axis: Total potential $U(x)$. Dashed horizontal
line indicates thermal energy above minimum. Voltages shown: 0.05 V
(a-b), 0.14 V (c-d), 0.29 V (e-f). $T=300$ K (left column) and $T=77$
K (right column), $\chi_{0}=0.0361$ eV. }
\label{Figure: J(x)} 
\end{figure*}

We begin by examining the effective potential energy experienced
by the mechanical degree of freedom 
\begin{equation}
U(x)=\frac{1}{2}m\omega_{0}^{2}x^{2}-\int_{x_{0}}^{x}dxF_{\text{e,neq}}(x),\label{U}
\end{equation}
where the electronic force $F_{\text{e,neq}}$ is given by {   Eq. }(\ref{fe})
and choice of $x_{0}$  is arbitrary.

The inclusion of the nonequilibrium electronically-induced force on
the $x$ coordinate is observed to introduce a double-well potential
feature to the ordinarily harmonic oscillator for some parameters.
Firstly, considering high temperature (300 K), it is observed that
for low voltages $V<0.11$ V the effective potential is roughly parabolic
with a single potential minimum. As the voltage is increased the effective
potential widens and a centralised energy barrier $U_{B}$ emerges
separating two minima. As can be seen in Figures \ref{Figure: J(x)}(a),(b),(c),
this change is gradual and initially the barrier is below the thermal
energy with respect to the lowest minima, $(U_{B}<k_{B}T)$, but for
higher voltages $V>0.17$ V it is above $(U_{B}>k_{B}T)$. This behaviour
continues up to $V=0.34$ V, after which the barrier begins to subside.
The qualitatively similar behaviour is observed at $T=77$ K, although
the low temperature ensures that the barrier, once present, always
exceeds thermal energy; shown in Figures \ref{Figure: J(x)}(b),(d),(f).
The influence of the nonequilibrium electronic force is more pronounced
at low temperature -- additional minima start to develop at smaller
voltage and the minima are separated by higher potential barriers.

\begin{figure}
\begin{centering}
\includegraphics[scale=0.6]{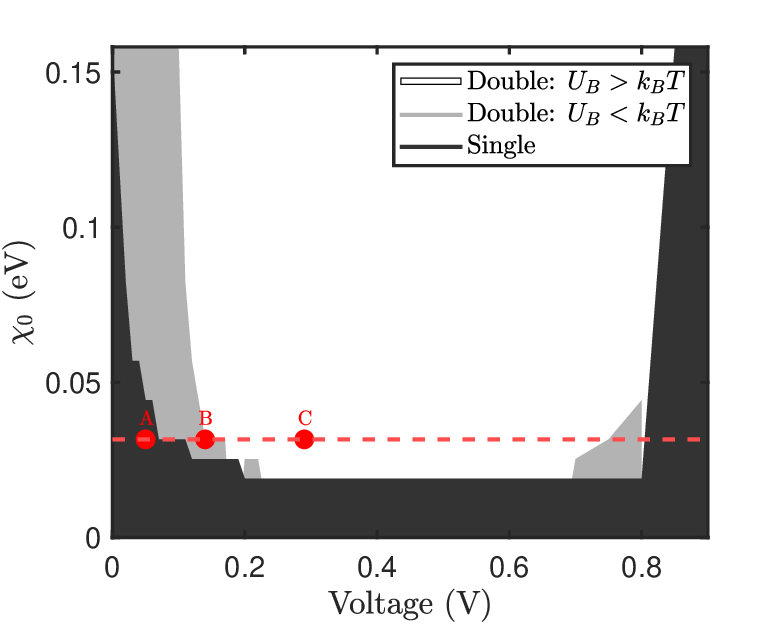} 
\par\end{centering}
\caption{System parameter map indicating the nature of the effective potential
$U(x)$ (\ref{U}) for the combination of electron-mechanical motion
strength $\chi_{0}$ and voltage $V$. The potential is defined by
the number of local minima: either a single or double potential well
and, for the double potential, whether the barrier between minima
is greater than the thermal energy. Point A, B, and C correspond to
voltages 0.05 V, 0.14 V, and 0.29 V, respectively. Dashed line represents
$\chi_{0}=0.0316$ eV. $T=300$ K.}

\label{Figure:PotentialMap} 
\end{figure}

The parameter space is further explored at $T=300$ K in Figure \ref{Figure:PotentialMap}
where the shaded regions indicate the qualitative nature of the potential
resulting from the strength of electron-mechanical-motion coupling
$\chi_{0}$ and voltage $V$. At the relatively low voltages considered,
the molecular electronic population tends to increase with voltage,
and so does the electronic force (\ref{fe}). This force is due to
the electron-mechanical motion interaction, so $\chi_{0}$ also scales
the magnitude and thus both parameters contribute to the development
of double-minima potential. In addition to the distinction between
single well and double well potentials, Figure \ref{Figure:PotentialMap}
also emphasises the distinction between the double-potential when
the barrier is above and below the thermal energy. Results presented
in \ref{Figure: J(x)}(a),(b), and (c) correspond to points A, B,
C in Figure \ref{Figure:PotentialMap}. We {    use $\chi_{0}=0.0316$
eV }in all our calculations for the rest of the paper.

\begin{figure}
\begin{centering}
\includegraphics[scale=0.6]{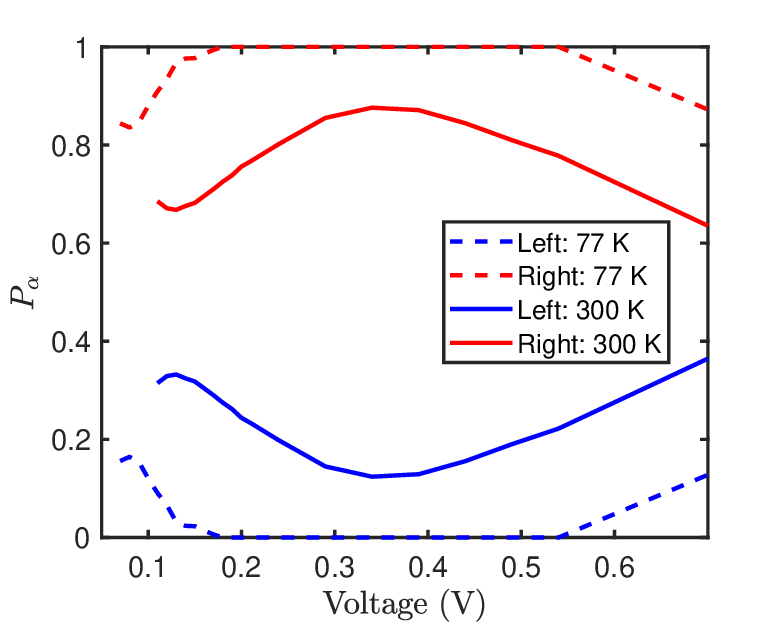} 
\par\end{centering}
\caption{Occupation probability for each left/right region of double potential
in blue/red, respectively, computed at temperatures $T=300$ K and
$T=77$ K.}

\label{Figure: OccupationWaiting} 
\end{figure}

To understand the dynamics of DNA's mechanical motion in a bistable
potential, it is instructive to view occupation probabilities for left
and right minima. For this, we perform Langevin simulations by numerically
solving {   Eqs.} (\ref{le1}, \ref{le2}) at $T=300$ K and $T=77$
K. These occupation probabilities are defined as $P_{\alpha}=\tau_{\alpha}/\tau_{\text{total}}$,
where $\tau_{\alpha}$ is the {   time which the mechanical coordinate
spends in potential side $\alpha=L,R$ during the total trajectory time $\tau_{\text{total}}$.}
As can be seen from the occupation of the left and right minima in
Figure~\ref{Figure: OccupationWaiting}, in the room temperature
regime ($T=300$ K), the occupation probability of the left minima
is not substantially suppressed even when the barrier exceeds the
thermal energy. Thermal fluctuations allow the coordinate to overcome
the barrier. In the lower temperature regime ($T=77$ K), the leftmost
minima is ``frozen out'' of the dynamics leading to an effectively
zero occupation when the barrier energy is sufficiently above the
temperature.

\subsection{Mechanical motion induces spin-polarisation of electronic current}

\begin{figure}
\begin{centering}
\includegraphics[scale=0.6]{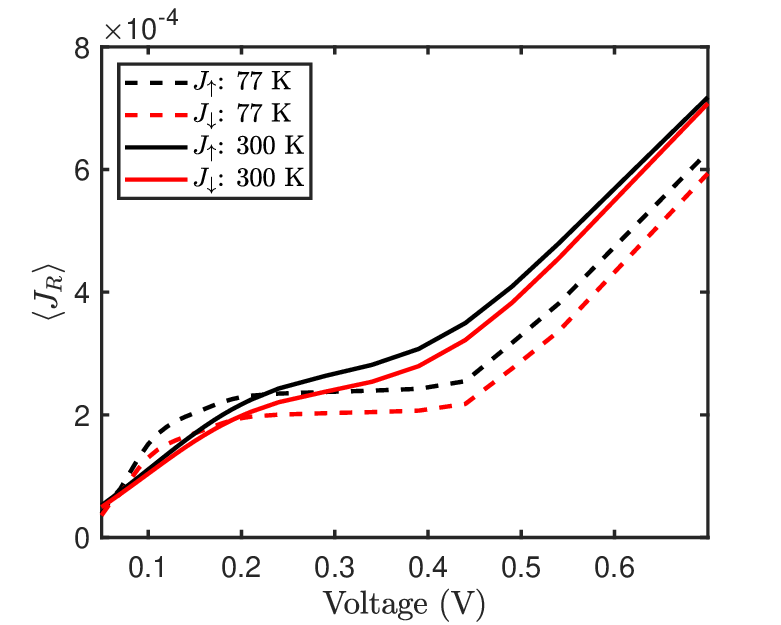} 
\par\end{centering}
\caption{Time-averaged spin currents computed at $T=300$ K and $T=77$ K as
a function of voltage. }
\label{Figure-2:-SpinCurrent} 
\end{figure}

\begin{figure}
\begin{centering}
\includegraphics[scale=0.6]{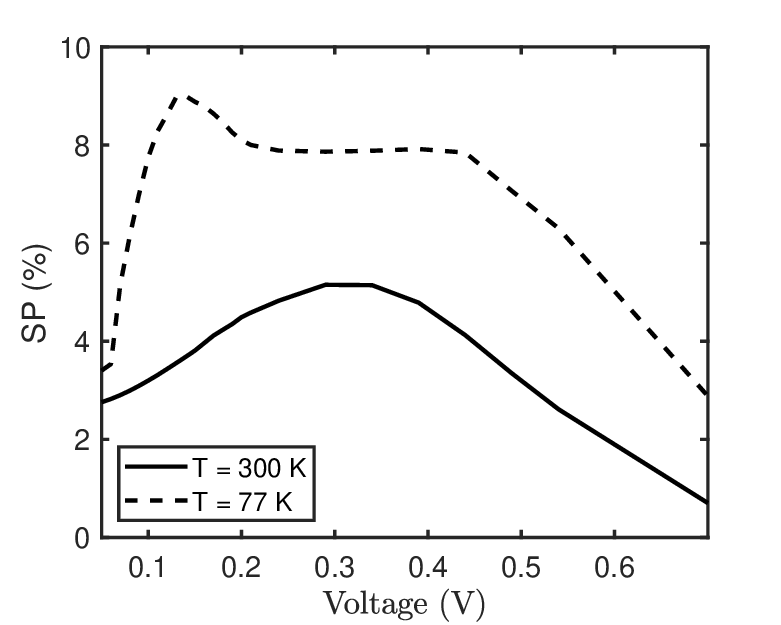} 
\par\end{centering}
\caption{Spin polarisation, $\text{SP}=(J_{\uparrow}-J_{\downarrow})/(J_{\uparrow}+J_{\downarrow})$,
computed at $T=300$ K and $T=77$ K as a function of voltage. }
\label{Figure-2:-SP} 
\end{figure}

Figure \ref{Figure: J(x)}(a)-(f) shows a nontrivial dependence of
the spin-resolved electric currents on the mechanical coordinate $x$.
The metastable left minima, which is populated more at higher temperatures,
corresponds to more conductive DNA junctions -- the electronic current
is larger for both spin components here in comparison to the stable right
minima. These differences in electronic currents for stable and metastable
DNA configurations are reflected in the current-voltage characteristics
shown in Figure \ref{Figure-2:-SpinCurrent}. As one can see in Figure
\ref{Figure-2:-SpinCurrent}, the electronic current computed at $T=300$
K is larger than the current at $T=77$ K; this observation is qualitatively
consistent with the experimental data \cite{kumar2022}. Note that
there is no noticeable temperature dependence of {    the electronic current
in the absence of coupling between the DNA's mechanical motion and the electronic degrees
of freedom.}

Figure \ref{Figure: J(x)}(a)-(f) also shows that the DNA's mechanical
motion in the stable (right) potential induces considerable majority-spin
polarisation of the current whereas metastable DNA configuration favours
minority-spin to a lesser extent. Given that high temperature enables
greater access to the left region, the time-averaged current will
receive greater contribution to minority-spin current. As a result,
the spin polarisation of electronic current decreases with increasing
temperature as one can see from Figure \ref{Figure-2:-SP}. For
$T=300$ K, spin-polarisation increases from roughly 2-5\% and is
maximised at 0.3 V. Beyond this the spin polarisation reduces with
voltage. Meanwhile for $T=77$ K, spin-polarisation increases from
3-9\% up to 0.2 V. SP maintains up to 0.4 V then also decays at high
voltage. This temperature dependence is contrary to the aforementioned
experiment \cite{kumar2022} -- we delegate this problem to future
study.

\subsection{Spin-resolved current noise assisted by DNA's mechanical motion}

Being a time-averaged quantity, the current-voltage characteristic
(Figure \ref{Figure-2:-SpinCurrent}) does not provide information
about the DNA dynamics; it does not reveal the important time-scales
related to the effective potential shapes. To reveal experimentally
accessible information about DNA's mechanical motion and its interplay
with spin polarisation, we turn our attention to spin-resolved current
noise calculations. Note also that the dynamical (velocity dependent)
corrections to the electric current do not play any role so far, since
their contributions disappear upon averaging over trajectory due to
being linear in the velocity. However, the corrections play a role
in the current noise.

Noise spectroscopy in molecular junctions has proven useful in identifying
transport mechanisms not discernible from the current-voltage characteristics
alone. Such measurements can reveal atomistic details of the local
environment and metal-molecule interfaces \cite{noise10,doi:10.1021/acs.nanolett.5b01270},
coupling between electronic and vibrational degrees of freedom \cite{galperin06a,PhysRevLett.106.136807,PhysRevLett.108.146602,thoss14,kosov17-nonren},
individual conduction transport channels \cite{doi:10.1021/nl060116e,doi:10.1021/nl904052r,Tsutsui:2010aa,noise17},
and mechanical stability of the junction \cite{C4NR03480E}. Naturally,
extending these ideas to consider spin noise provides insight into
spin transport. Indeed, shot noise has been used to identify spin-polarised
transport within a specific spin channel \cite{burtzlaff2015}. Spin
fluctuations can reveal Rabi splittings, Zeeman shifts, and the formation
of doublet and triplet states \cite{glasenapp2014}; also enabling
disturbance-free probing of spin dynamics \cite{oestreich2005,crooker2004}.
Shot noise calculations \cite{mirzanian2013,pradhan2018} have explored
contributions to quantum noise and its correlations; however the role
of molecular, mechanical motion is largely unexplored. In what follows,
we quantify the noise as a result of this motion.

The spin-resolved current noise is formally defined as 
\begin{equation}
S_{\alpha,\sigma\sigma'}(\tau)=\lim_{T\rightarrow+\infty}\frac{1}{T}\int_{0}^{T}dt\langle\Big[\delta\hat{J}_{\alpha\sigma}(t),\delta\hat{J}_{\alpha\sigma}(t+\tau)\Big]_{+}\rangle,\label{s1}
\end{equation}
where $\delta\hat{J}_{\alpha\sigma}(t)$ describes the instantaneous
deviation of the $\sigma$-spin current at time $t$ from its mean
value and $[...,...]_{+}$ is the anti-commutator. The quantum expectation
value $\langle...\rangle$ averages over electronic degrees of freedom
while the time average over the mechanical motion of DNA is described
by $\lim_{T\rightarrow+\infty}\frac{1}{T}\int_{0}^{T}dt...$; equivalent
to an ensemble average of DNA geometries. This total noise is a complex
amalgamation of various sources of noise. Quantum noise is inherent
to electrons due to discrete charge, Pauli exclusion principle, shot
noise, finite temperature and the correlations arising from electron-electron
interactions. The latter is excluded from our model of non-interacting
electrons. Additionally, a mechanical noise results from the current-induced
motion of the DNA. Due to the time-scale separation of electrons and
this molecular motion, this contribution from the noise can be isolated
\cite{preston2020}. The characteristic time scale of shot noise decay
is $1/\Gamma$, whereas the noise due to nuclear motion appears on
much longer times, $1/\zeta$. For the viscosity considered, $\zeta/\Gamma\approx0.006$;
hence the noise induced by geometrical fluctuations dominates the
noise power spectrum at low frequencies, and can exceed the shot noise
contribution by orders of magnitude. In what follows we focus on the
``mechanical'' noise as 
\begin{equation}
S_{\alpha,\sigma\sigma'}(\tau)=2\lim_{T\rightarrow+\infty}\frac{1}{T}\int_{0}^{T}dt\delta J_{\alpha\sigma}(t+\tau)\delta J_{\alpha\sigma'}(t),\label{currentnoise}
\end{equation}
where the current fluctuation at time $t$ is 
\begin{equation}
\delta J_{\alpha\sigma}(t)=J_{\alpha\sigma}(t)-\lim_{T\rightarrow+\infty}\frac{1}{T}\int_{0}^{T}dtJ_{\alpha\sigma}(t).
\end{equation}

The adiabatic mechanical noise 
\begin{equation}
S_{\alpha,\sigma\sigma'}^{(0)}(\tau)=2\lim_{T\rightarrow+\infty}\frac{1}{T}\int_{0}^{T}dt\delta J_{\alpha\sigma}^{(0)}(t+\tau)\delta J_{\alpha\sigma'}^{(0)}(t),\label{S0}
\end{equation}
and dynamical corrections, which not only depend on the instantaneous
position but also on the velocity, 
\begin{multline}
S_{\alpha,\sigma\sigma'}^{(1)}(\tau)=2\lim_{T\rightarrow+\infty}\frac{1}{T}\int_{0}^{T}dt\\
\Big[\dot{x}(t+\tau)B_{\alpha\sigma}(t+\tau)\delta J_{\alpha\sigma'}^{(0)}(t)+\delta J_{\alpha\sigma}^{(0)}(t+\tau)\dot{x}(t)B_{\alpha\sigma'}(t)\\
+\dot{x}(t+\tau)B_{\alpha\sigma}(t+\tau)\dot{x}(t)B_{\alpha\sigma'}(t)\Big],\label{S1}
\end{multline}
are included in the total noise 
\begin{equation}
S_{\alpha,\sigma\sigma'}(\tau)=S_{\alpha,\sigma\sigma'}^{(0)}(\tau)+S_{\alpha,\sigma\sigma'}^{(1)}(\tau).\label{S}
\end{equation}
Here we factorised the dynamical correction to the electric current 
\begin{equation}
J_{\alpha\sigma}^{(1)}(t)=\dot{x}B_{\alpha\sigma}(x).
\end{equation}
As we will demonstrate, the terms (\ref{S0}), (\ref{S1}) are responsible
for different features of noise correlation functions. The adiabatic
mechanical noise $S_{\alpha,\sigma\sigma'}^{(0)}(\tau)$ is generally
featureless and shows overall exponential decay. The first two cross
correlation terms in $S_{\alpha,\sigma\sigma'}^{(1)}(\tau)$ (the
terms which are linear in velocity of mechanical motion) are responsible
for the deviation of $S_{\alpha,\uparrow\downarrow}(\tau)$ from $S_{\alpha,\downarrow\uparrow}(\tau)$
for bistable mechanical motion. The last term in $S_{\alpha,\sigma\sigma'}^{(1)}(\tau)$
involves velocity-velocity correlations containing the underlying
dip and peak features at half-period and full-period of motion in
the right region.

\begin{figure}
\begin{centering}
\includegraphics[scale=0.6]{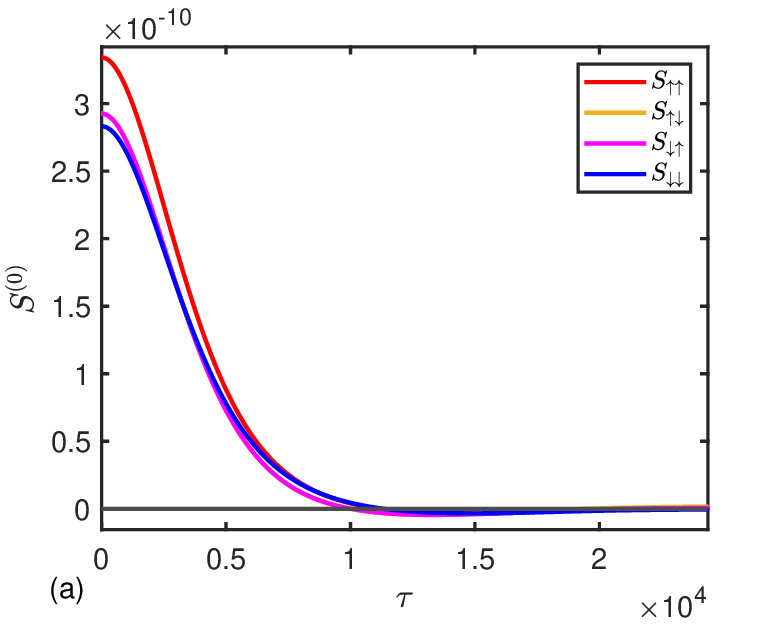} 
\par\end{centering}
\begin{centering}
\includegraphics[scale=0.6]{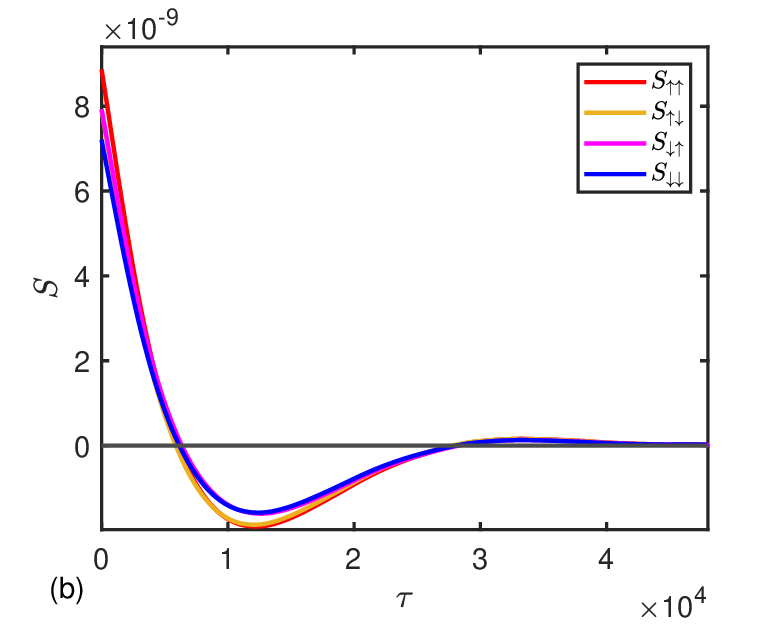} 
\par\end{centering}
\caption{Spin-resolved noise $S_{R,\sigma\sigma'}(\tau)$ for single-minima
effective potential (corresponds to point A in Figure \ref{Figure:PotentialMap}).
(a) Adiabatic noise (\ref{S0}), (b) total noise computed with dynamical
corrections (\ref{S}).}
\label{Figure:-Covariance_RegionA} 
\end{figure}

Figures \ref{Figure:-Covariance_RegionA} (a,b) show spin resolved
components of current noise for DNA's mechanical motion in single-potential
shown in Figure \ref{Figure: J(x)}-a. Adiabatic mechanical noise
$S^{(0)}(\tau)$ demonstrates simple decay of correlations in time
with no distinct features. Once dynamical corrections are included, the
noise reveals negative correlations at times comparable to the half-period
- this dip is a characteristic feature of classical velocity-velocity
correlation functions and, unsurprisingly, it can be shown that it
originates predominantly from the last term in {   Eq.} (\ref{S1}) which contains
$\dot{x}(t+\tau)\dot{x}(t)$.

\begin{figure}
\begin{centering}
\includegraphics[scale=0.6]{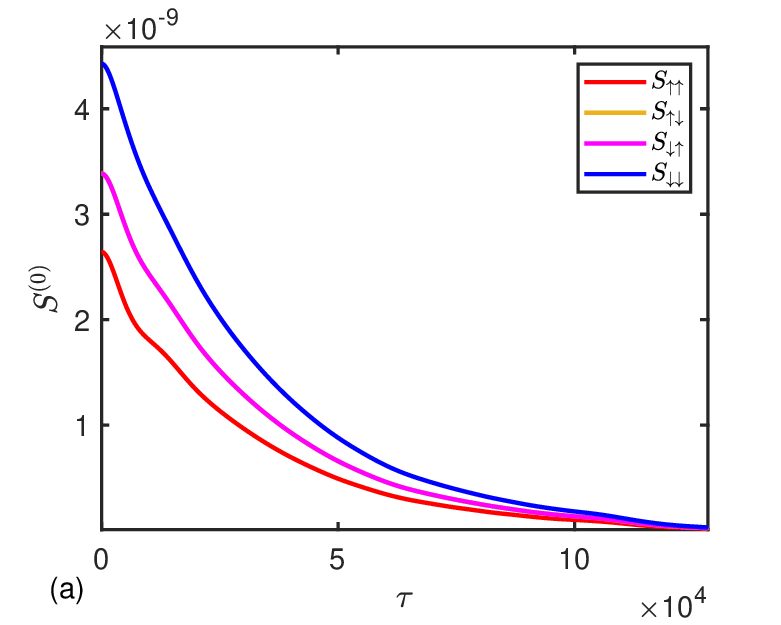} 
\par\end{centering}
\begin{centering}
\includegraphics[scale=0.6]{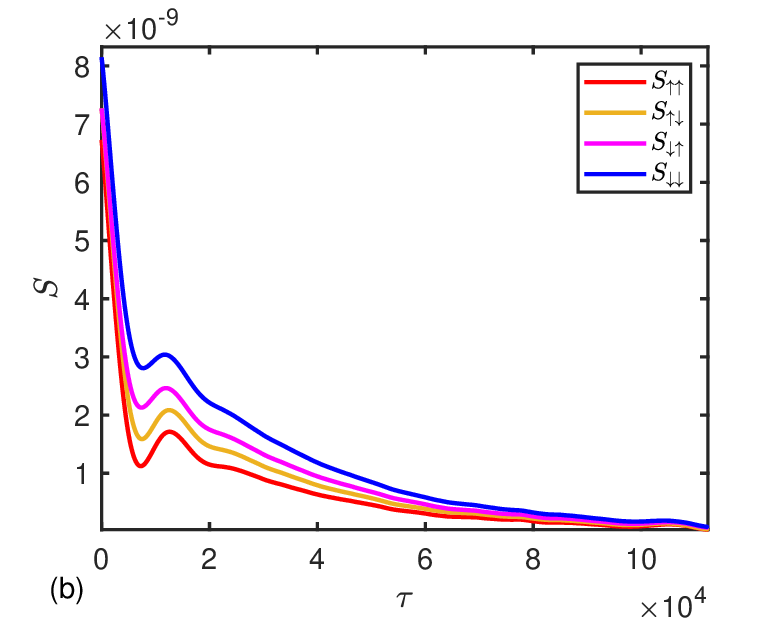} 
\par\end{centering}
\caption{Spin-resolved noise $S_{R,\sigma\sigma'}(\tau)$ for bistable mechanical
motion (corresponds to point C in Figure \ref{Figure:PotentialMap}).
a) Adiabatic noise (\ref{S0}), (b) total noise computed with dynamical
corrections (\ref{S}). }
\label{Figure:-Covariance_RegionD} 
\end{figure}

Figures \ref{Figure:-Covariance_RegionD} (a,b) show the spin resolved
components of current noise for DNA bistable mechanical motion shown
in Figure \ref{Figure: J(x)}-e. Adiabatic noise $S^{(0)}(\tau)$
remains always positive and begins to develop small positive-correlation
features, due to the consistently large deviation of the current from
its average value for stable (right) potential and at the same time
consistently smaller current dispersion in metastable (left) potential
- see \ref{Figure: J(x)}-e. Dynamically corrected noise reveals a
distinct negative contribution to correlations (dip) at approximately the
half-period of mechanical motion in the stable potential and a positive-correlation
(peak) appears at approximately of full period of above motion - this
feature again takes it origin from the velocity-velocity correlation
term in {   Eq.} (\ref{S1}). It is interesting to note that although the average
current is dominated by $\sigma=\uparrow$ electrons - see Figure
\ref{Figure-2:-SP}, the noise shows the opposite spin selectivity:
$S_{\alpha,\uparrow\uparrow}(\tau)$ is always smaller than $S_{\alpha,\downarrow\downarrow}(\tau)$
due to smaller deviation of $J_{\alpha\uparrow}(x)$ from the mean
compared to $J_{\alpha\downarrow}(x)$. 

The interesting new feature associated with mechanical bistability
is the different temporal correlations between fluctuations of spin-up
and spin-down currents depending upon which fluctuations occurred
first -- as it is apparent from Figure \ref{Figure:-Covariance_RegionD}
(b), cross-spin noise $S_{\alpha,\uparrow\downarrow}(\tau)$ deviates
from $S_{\alpha,\downarrow\uparrow}(\tau)$. Analysis of the sum of
two terms in current noise dynamical corrections, $\dot{x}(t+\tau)B_{\alpha\sigma}(t+\tau)\delta J_{\alpha\sigma'}^{(0)}(t)$
and $\delta J_{\alpha\sigma}^{(0)}(t+\tau)\dot{x}(t)B_{\alpha\sigma'}(t)$
shows that the up-down component is negative whereas the down-up component
is positive (at the timescale of noise correlation decay).

\section{Conclusions}

We have explored mechanical motion due to current induced forces in
a double stranded DNA helix with spin-orbit-coupling and its role
in the spin polarisation of current and noise induced by DNA chirality.
We represent the mechanical motion {   according to the} dynamics of a stochastic classical
variable experiencing fluctuations and dissipation induced by the
environment as well as forces exerted by nonequilibrium, current carrying
electrons. The electronic degrees of freedom are described quantum mechanically
using NEGF. NEGF are computed along the trajectory for the classical
variable taking into account dynamical, velocity dependent corrections.

We observe that 
\begin{itemize}
\item DNA's mechanical instability is induced by tunneling electrons. The
instability emerges at moderate applied voltage bias given that the
strength of electron-mechanical motion coupling exceeds a certain critical
value. The physical regimes for stable and bistable DNA motion are
identified. 
\item Mechanical motion results in moderate increase of SP, achieving 3-9\%
depending on the temperature and voltage bias.This
is compared to the absence of any SP without mechanical motion. The
comparatively small SP is expected given the short helix. 
\item The temperature-dependent energy landscape together with temperature-dependent
dynamics of the mechanical motion lead to considerable temperature
dependence of the electronic current. The temperature dependence of
the total current agrees with experimental observation, however, SP
does not show the expected response to increased temperature. This
raises questions about the importance of additional interactions such
as allowing the mechanical motion to flip the spin via SOC mechanisms
which are not included in the present study. 
\item The spin resolved noise induced by DNA's mechanical motion can be
a useful experimental tool to extract information about DNA mehanical
motion and detect the emergence of mechanical instabilities. 
\end{itemize}
\bibliographystyle{ieeetr}

\end{document}